\documentclass[12pt]{iopart}

\usepackage{graphicx}
\usepackage{siunitx}
\DeclareSIUnit \belm {bm}
\DeclareSIUnit{\dBm}{\deci\belm}

\usepackage[colorlinks=true, linkcolor=blue,citecolor=red]{hyperref}
\usepackage{xcolor}
\usepackage{ulem}

\newcommand{\ii}{{\rm i}}
\newcommand{\ee}{{\rm e}}

\newcommand{\be}{\begin{equaion}}
\newcommand{\bea}{\begin{eqnarray}}
\newcommand{\eea}{\end{eqnarray}}

\definecolor{green}{rgb}{0.1,.7,0.05}

\newcommand{\calP}{{\mathcal P}}
\newcommand{\calC}{{\mathcal C}}
\newcommand{\calR}{{\mathcal R} }
\newcommand{\calT}{{\mathcal T}}
\newcommand{\calS}{{\mathcal S}}
\newcommand{\calF}{{\mathcal F}}

\newcommand{\calV}{{\mathcal V}}

\newcommand{\calL}{{\mathcal W}}

\newcommand{\calN}{{\mathcal N}}

\begin{document}

\title{Two--membrane cavity optomechanics: non--linear dynamics}

\author{
Paolo~Piergentili$^{1,2}$, Wenlin~Li$^{1}$, Riccardo~Natali$^{1,2}$, Nicola~Malossi$^{1,2}$, David~Vitali$^{1,2,3}$, Giovanni~Di~Giuseppe$^{1,2}$}
\address{	$^1$ \text{School of Science and Technology, Physics Division, University of Camerino, }\\
		\quad\text{I-62032 Camerino (MC), Italy}\\
		$^2$ \text{INFN, Sezione di Perugia, Italy}\\
		$^3$ \text{CNR-INO, L.go Enrico Fermi 6, I-50125 Firenze, Italy}
		}
\ead{gianni.digiuseppe@unicam.it, david.vitali@unicam.it}

\vspace{10pt}
\begin{indented}
\item[\today]
\end{indented}

\begin{abstract}
We study the non--linear dynamics of a multimode optomechanical system constituted of a driven high--finesse Fabry--P\'erot cavity containing two vibrating dielectric membranes.
The analytical study allows to derive a full and consistent description of the displacement detection by a probe beam in the non--linear regime, enabling the faithful detection of membrane displacements well above the usual sensing limit corresponding to the cavity linewidth.
In the weak driving regime where the system is in a pre--synchronized situation, the unexcited oscillator has a small, synchronized component at the frequency of the excited one; both large and small amplitude resonator motions are transduced in a nontrivial way by the non--linear response of the optical probe beam. 
We find perfect agreement between the experimental results, the numerical simulations, and an analytical approach based on slowly--varying amplitude equations. 
\end{abstract}

\date{\today}

\section{Introduction}
Multimode optomechanical systems~\cite{Aspelmeyer2014} are attracting an increasing interest for the study of collective dynamical effects, both at quantum and classical level. Two different situations are mainly considered from both the theoretical and experimental point of view: i) a group of mechanical oscillators interacting via radiation pressure with the \textit{same} optical mode~\cite{Bhattacharya:2008ab,Hartmann:2008aa,Xuereb:2012fk,Holmes:2012aa,tomadin,Xuereb:2013ys,seok,Xuereb3,Li:2016aa,Nair:2016aa,Li:2017ac,Weaver:2017aa,Piergentili:2018aa,
Gartner2018,Wei2019,Naesby2019} (e.g. multiple membranes within the same optical cavity); ii) an array of mechanical oscillators each interacting locally with a single optical mode, and coupled by the tunneling of photons and phonons between neighboring sites~\cite{Chang,Heinrich:2011ab,Ludwig:2013aa,Chen,Schmidt,Peano}, (e.g. optomechanical crystals in one and two dimensions~\cite{Eichenfield}).

Several features of multimode optomechanical systems have already been investigated in the literature, such as long-range collective interactions~\cite{Xuereb:2012fk,Xuereb:2013ys,Xuereb3} yielding an effective increase of the optomechanical coupling, slowing and stopping light~\cite{Chang}, correlated quantum many-body states~\cite{Ludwig:2013aa}, reservoir engineering and dynamical phase transitions~\cite{tomadin}, graphene--like Dirac physics~\cite{Schmidt}, topological phases of sound and light~\cite{Peano}, transport in a one-dimensional chain~\cite{Chen,Gan,Xiong}, superradiance and collective gain~\cite{Kipf}, and nonreciprocal routing of electromagnetic signals~\cite{Bernier:2017,Peterson:2017}. 

%
The radiation pressure interaction is inherently non-linear and the effects of such non-linearity on the mechanical motion are easily manifested when the optical cavity is driven on the blue sideband, when optical backaction is responsible for mechanical antidamping~\cite{Aspelmeyer2014}. When the latter overcomes
the internal mechanical friction, a Hopf bifurcation towards a regime of self-induced mechanical oscillations
takes place~\cite{Carmon,Kippenberg2005,Marquardt2006,Metzger,Krause:2015aa,Buks2019,Piergentili:2020a}, with a fixed amplitude, and a free running oscillation phase, which may lock to external forces or to other optomechanical oscillators~\cite{Balanov2008}. This mutual phase-locking of self-oscillating resonators is at the basis of optomechanical synchronization, which has been thoroughly investigated both theoretically~\cite{Holmes:2012aa,Heinrich:2011ab,Ludwig:2013aa,Mari2013,Ying2014,Wang2014,Weiss2016,Li2016,Bemani2017,Li2017,Li2020}, and experimentally~\cite{Zhang:2012aa,Bagheri:2013ht,Agrawal:2013fk,Matheny:2014fv,Shah2015,Zhang:2015ad,Huang2017,GilSantos2017,Colombano2019,Sheng:2020aa} under different configurations. The non-linear effects of radiation pressure manifest themselves whenever the mechanical motion produces a cavity frequency shift comparable or larger than
the optical linewidth, resulting in a nontrivial modification of the cavity response to the external driving. This is responsible for a variety of non-linear phenomena beyond synchronization, such as phonon lasing~\cite{Grudinin2010}, mode competition~\cite{Kemiktarak:2014cq}, and chaos~\cite{Carmon2007,Navarro2017,Wu2017}. This radiation-pressure-induced non-linear behavior may occur not only when the mechanical resonators are driven to large amplitude via the
parametric amplification provided by blue-sideband driving, but also in the strong optomechanical coupling regime~\cite{Leijssen:2015aa} where even intrinsic Brownian
motion induces cavity frequency fluctuations larger than the optical linewidth~\cite{Leijssen:2017aa,Cattiaux2020}. In both situations, the optomechanical non-linearity plays a fundamental role, affecting optomechanical displacement measurement and transduction, and this role can be exploited for extending in a nontrivial way the dynamic range of optomechanical sensors beyond the cavity linewidth regime~\cite{Javid2020}.

Here we experimentally explore the non-linear dynamics of the multimode optomechanical setup first demonstrated in Ref.~\cite{Piergentili:2018aa}, realized by placing a membrane cavity within a high-finesse Fabry--P\'erot cavity. Ref.~\cite{Piergentili:2018aa} reported a $\sim 2.47$ gain in the optomechanical coupling strength of the membrane relative motion with respect to the single membrane case, and showed the capability to tune the single-photon
optomechanical coupling on demand. Ref.~\cite{Sheng:2020aa} recently demonstrated synchronization of this two-membrane cavity optomechanical system, by operating with a low-finesse cavity in the strongly unresolved sideband regime.
Here instead we focus onto the pre-synchronization regime of weak blue-detuned driving, where only one of the two membrane resonators enters into a limit cycle through the Hopf bifurcation, while the other resonator remains in a mixed condition where the modulation of the radiation pressure force induced by the excited oscillator does not prevail over the thermal motion. We provide a detailed, \textit{quantitative} analysis of the dynamics in this regime, with a significant agreement between the experimental data, the numerical simulation, and the analytical treatment based on amplitude equations of Ref.~\cite{Li2020}. This quantitative analysis is based on a detailed treatment of the optical detection apparatus including the probe and calibration tones, and provides an accurate, reliable, measurement of the displacement of both membranes, even in the non-linear regime where the frequency modulation caused by the two membranes' motion is significantly larger than the cavity linewidth. A remarkable result of this analysis is that, in the presence of a self-oscillating resonator in a limit cycle, non-linear corrections to the displacement measurement by the probe cavity output must be applied not only to the excited resonator but also to the small-amplitude, unexcited one. This implies that in multimode optomechanical systems, whenever multiple  mechanical resonators are detected by the same single probe field (such as for example in Refs.~\cite{Zhang:2012aa,Bagheri:2013ht,Shah2015,Zhang:2015ad,GilSantos2017}), and at least one resonator enters a limit cycle, one has to properly include the full non-linear dynamics of the system in order to extract the correct displacement measurement from the output probe spectrum.

The paper is organized as follows: In Sec.~2 we provide the basic theoretical description of the multimode optomechanical system under study. In Sec.~3 we describe the experimental setup, and in Sec.~4 we derive in detail the probe beam power spectral density, including all the non-linear effects. In Sec.~5 we analyze the non-linear dynamics of the mechanical modes at the onset of synchronization and we provide an analytical description in very good agreement with the numerical and experimental results. Sec.~6 is for concluding remarks.

\section{Theoretical description of the system dynamics}
\label{System dynamics}

We study the non-linear dynamics of a multimode optomechanical system, formed by two electromagnetic and two mechanical modes, at room temperature, which justifies a treatment in terms of {\it classical} amplitudes, and implies that thermal noise will be dominant for the mechanical modes and treated as {\it classical} complex random noises.
The two optical modes with frequencies $\omega_{ci}$, $i=(1,2)$, total cavity amplitude decay rates $\kappa_i=\kappa_{in,i}+\kappa_{ex,i}$ with $\kappa_{ex,i}$ optical loss rates through all the ports different from the input one $\kappa_{in,i}$, and driven at frequencies $\omega_{Li}$, interact via radiation--pressure with two mechanical modes with resonance frequencies $\omega_j$, $j=(1,2)$, mass $m_j$, and amplitude decay rates $\gamma_j$. Their dynamics is described by the set of coupled classical Langevin equations for the corresponding optical and mechanical complex amplitudes $\alpha_i(t)$ and $\beta_j(t)$~\cite{Wang2014,Weiss2016,Li2017}, respectively,
\begin{eqnarray}\label{eq:c_langevin1}
	\dot{\alpha}_i(t)=&\left(\ii \Delta^{(0)}_i-\kappa_i\right)\alpha_i(t) \!+\! E_i \!
			+\! \sum_{j=1,2} 2\rm{i}g_{ij}\text{Re}[\beta_j(t)]\alpha_i(t)
		\!+\!\sqrt{2\kappa_i}\,\alpha_i^{opt}(t),\\
	\dot{\beta}_j(t)=&(-\ii \omega_j-\gamma_j)\beta_j(t)+\ii\sum_{i=1,2}g_{ij}\vert\alpha_i(t)\vert^2
			+\sqrt{2\gamma_j}\,\beta^{in}_j(t)\,,
\label{eq:c_langevin}
\end{eqnarray}
where $\Delta^{(0)}_i=\omega_{Li}-\omega_{ci}$ are the detunings, $E_i=\sqrt{2\kappa_{in,i} P_i/\hbar\omega_{Li}}$ the driving rates with $P_i$ the associated laser input powers,  $g_{ij}=-(d\omega_{c i}/dx_j)x_{{\rm zpf},j}$ the single-photon optomechanical coupling rates, $x_{{\rm zpf},j}=\sqrt{\hbar/2m_j\omega_j}$ the spatial width of the $j$-th oscillator zero point motion, and $\beta^{in}_j(t)$, and $\alpha_i^{opt}(t)$, are the mechanical and optical noise terms, respectively.
These noises are uncorrelated from each other and the only nonzero correlation functions are $\langle \beta^{in,*}_j(t) \beta^{in}_{j'}(t')\rangle =(\bar{n}_j+1/2) \delta_{jj'}\delta(t-t')$, where $\bar{n}_j =\left[\exp\left(\hbar \omega_j/k_B T\right)-1\right]^{-1} \simeq k_B T/\hbar \omega_j \gg 1$ is the mean thermal occupation number.
Multimode optomechanical systems formed by two electromagnetic modes and two mechanical modes have been proposed and demonstrated~\cite{Bernier:2017,Peterson:2017} for the nonreciprocal routing of signals controlled by the relative phase of multiple external and off-resonant drives. In the present case the weak, quasi-resonant driving probe beam is used only for detecting the mechanical motion and we are far from the regime where one can use and control the relative phase of the drivings for nonreciprocal effects.
In this system, under appropriate parameter regimes, the pump cavity mode ($i=1$) may drive the oscillators into a self-sustained limit cycle~\cite{Carmon,Kippenberg2005,Marquardt2006,Metzger,Krause:2015aa,Buks2019}, which may eventually become synchronized. Synchronization may occur on a long timescale, determined by the inverse of the typically small parameters $\Delta \omega = \omega_2-\omega_1$ (typically never larger than few kHz), and $\gamma_j$ (order of Hz). Therefore it is physically useful to derive from the full dynamics of the classical Langevin equations (\ref{eq:c_langevin1})-(\ref{eq:c_langevin}), approximate equations able to correctly describe the slow, long--time dynamics of the two mechanical resonators, leading eventually to synchronization.

We adapt here the slowly varying amplitude equations approach of Ref.~\cite{Holmes:2012aa} to the case with noise studied here, as discussed in detail in Ref.~\cite{Li2020}, Discarding here the limiting case of chaotic motion of the two resonators, which however occurs only at extremely large driving powers, and are not physically meaningful for the Fabry--P\'erot cavity system considered here, it is known that each mechanical resonator, after an initial transient regime, sets itself into a dynamics of the following form
\begin{equation}\label{ansatz}
\beta_j(t)=\beta_{0,j}+A_j(t)\ee^{-\ii \omega_{\rm ref} t},
\end{equation}
where $\beta_{0,j}$ is the approximately constant, static shift of the $j-$th resonator, $A_j(t)$ is the corresponding slowly-varying complex amplitudes, and $\omega_{\rm ref} \gg \Delta \omega$ is a reference mechanical frequency, of the order of $\omega_j$.

From eqs.~(\ref{eq:c_langevin1})--(\ref{eq:c_langevin}) one gets the set of coupled amplitude equations~\cite{Li2020}
\begin{eqnarray}
	\dot{A}_1(t)=\left[-\gamma_1-\ii\Delta\omega_1\right] A_1(t)&+&\ii d_1A_1(t) +\ii d_{12} A_2(t) \nonumber \\
		&+&\ii\sum_{i=1,2}g_{i1}\eta_i^{opt}(t) + \sqrt{2\gamma_1}\beta^{in}_1(t), \label{eq:first-order finnal0}\\
	\dot{A}_2(t)=\left[-\gamma_2-i\Delta\omega_2\right] A_2(t)&+&\ii d_2A_2(t) +\ii d_{12} A_1(t) \nonumber \\
		&+&\ii\sum_{i=1,2}g_{i2}\eta_i^{opt}(t) + \sqrt{2\gamma_2}\beta^{in}_2(t),
\label{eq:first-order finnal1}
\end{eqnarray}
with $\Delta\omega_j = \omega_j - \omega_{\rm ref}$, and where
\begin{eqnarray}\label{bah1}
 &&d_1= \left(\frac{g_{11}^2 \mathcal{F}_1}{g_1^b}+\frac{g_{21}^2 \mathcal{F}_2}{g_2^b}\right), \\
 &&d_2= \left(\frac{g_{12}^2 \mathcal{F}_1}{g_1^b}+\frac{g_{22}^2 \mathcal{F}_2}{g_2^b}\right), \\
 &&d_{12}= \left(\frac{g_{11}g_{12} \mathcal{F}_1}{g_1^b}+\frac{g_{21}g_{22} \mathcal{F}_2}{g_2^b}\right),
\end{eqnarray}
are non--linear coefficients because of their dependence upon the regular dimensionless auxiliary functions $\mathcal{F}_i$, which are given by
\begin{equation}
	\mathcal{F}_i=\frac{E_i^2}{\vert A_i^b\vert}\sum_{n=-\infty}^{\infty}
	\frac{J_n\left(-\xi_i\right)J_{n+1}\left(-\xi_i\right)}
		{[\ii n\omega_{\rm ref}-\calL_i][-\ii (n+1)\omega_{\rm ref}-\calL_i^*]}\,,
\label{eq:auxiliary function}
\end{equation}
and which can be easily shown to be a function of even powers of $|A_i^b|$ only.
We have defined the {\it bright} complex amplitudes $A_i^b(t)=\vert A_i^b(t)\vert \ee^{\ii \theta_i(t)}=\sum_j g_{ij}A_j(t)/g_i^b$, $g_i^b=\sqrt{g_{i1}^2+g_{i2}^2}$, $\xi_i=2g_i^b\vert A_i^b\vert /\omega_{\rm ref}$, $\calL_i=\ii  \left[\Delta^{(0)}_i+\sum g_{ij}(\beta_{0,j}+\beta^{*}_{0,j})\right]-\kappa_i$, and $J_n$ is the $n$-th Bessel function of the first kind. $\eta_i^{opt}$ is the term describing the noise of optical origin~\cite{Li2020}.
As already shown in Refs.~\cite{Holmes:2012aa,Li2020}, eqs.~(\ref{eq:first-order finnal0})-(\ref{eq:first-order finnal1}) provide a general and very accurate description of the dynamics of the two mechanical resonators.
%

\section{Experimental setup}
\label{sec:experimental_setup}
\begin{figure}[b!]
\begin{center}
   {\includegraphics[width=.695\textwidth]{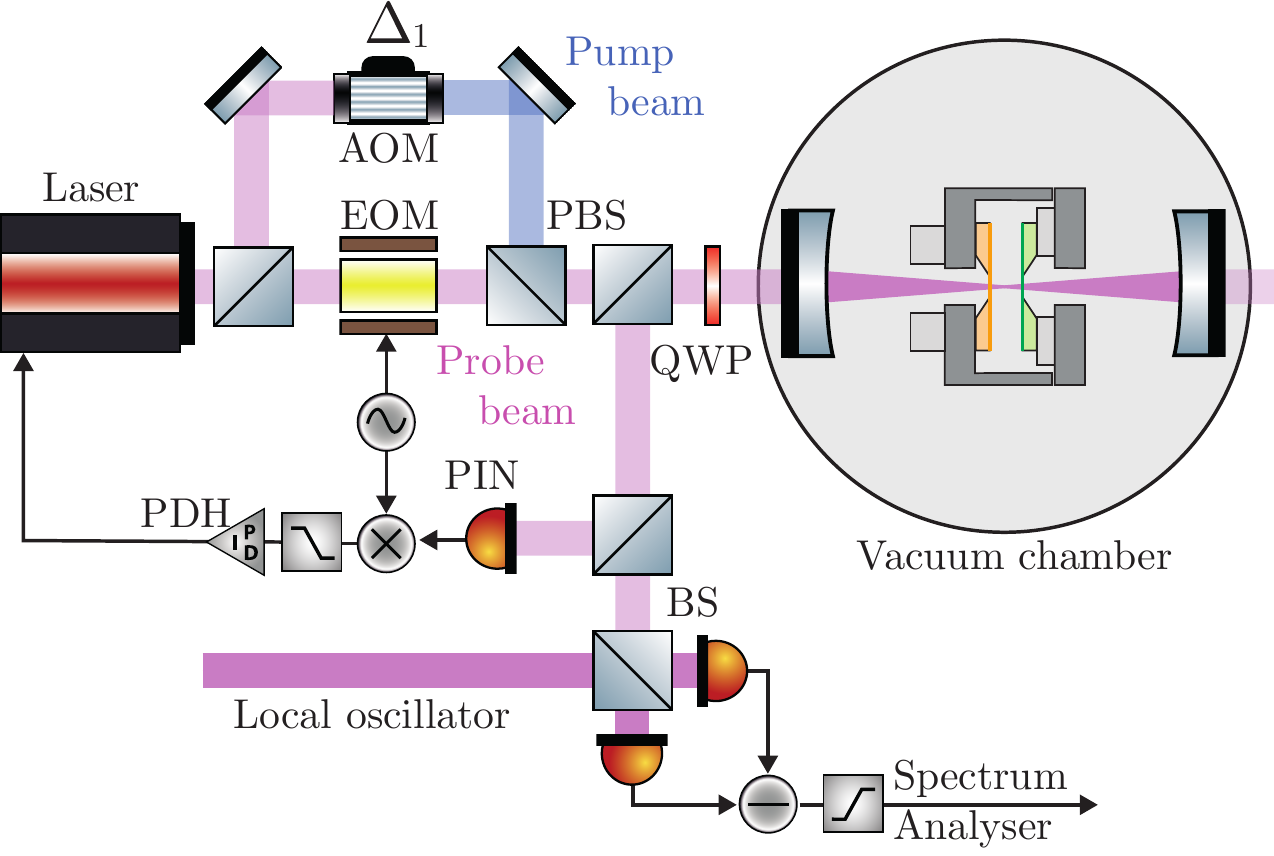}}
 \caption{
	Experimental setup for studying the non--linear dynamics in an optomechanical system constituted of a two-membrane sandwich within a cavity.
	A probe beam, frequency modulated by an electro-optical modulator (EOM), impinges on the optical cavity. The reflected beam is split: one component is detected, demodulated and low-pass amplified for generating the Pound--Drever--Hall (PDH) error signal able to lock the laser to the cavity; the second component is analyzed by homodyne detection in order to detect the mechanical motion. A further beam, the pump beam, detuned by $\Delta_1$ from the cavity resonance by means of an acousto--optic modulator (AOM), is turned on for engineering the optomechanical interaction, and in particular to realize laser driving of the mechanical modes. HWP denotes a half--waveplate, QWP a quarter--waveplate, BS a beam--splitter, and PBS a polarizing beam--splitter.
}
\label{fig:Figure_Setup}
\end{center}
\end{figure}
The experimental setup for studying the non-linear dynamics in an optomechanical system constituted of a two-membrane sandwich within a cavity, is shown in Fig.~\ref{fig:Figure_Setup}.
A laser beam at wavelength $\lambda_0 = \SI{1064}{\nano\meter}$ is split in a probe beam with intensity $P_{probe} =\SI{5.9}{\micro\watt}$, modulated by an electro-optical modulator (EOM), and a pump beam, detuned by $\Delta_1$ from the cavity resonance by means of an acousto-optic modulator (AOM). The reflected probe beam is locked to the optical cavity by means of a Pound--Drever--Hall (PDH) technique, and the thermal voltage spectral noise (VSN) is measured by homodyne detection of the light reflected by the optical cavity. The pump beam is used for engineering the optomechanical interaction, and in particular to realise laser driving of the mechanical modes.

The optical and mechanical properties of the optomechanical system were investigated in Ref.~\cite{Piergentili:2018aa}.
The membrane--cavity length, realised with two equal membranes (Norcada), was measured to be $L_{\rm c} = \SI{53.571\pm0.009}{\micro\meter}$, and the membrane thickness is $L_\mathrm{m} = \SI{106\pm1}{\nano\meter}$ that is found assuming the index of refraction of Si$_3$N$_4$ given in Ref.~\cite{Luke:2015aa}. Assuming rectangular membranes, and the nominal values provided by the manufacturer for the stress, $\sigma = \SI{.825}{\giga\pascal}$, and for the density $\rho = \SI{3100}{\kilo\gram\per\meter^3}$, the side lengths were estimated to be
$L_{x}^{(1)} = \SI{1.519 \pm 0.006}{\milli\meter}$, $L_{y}^{(1)} = \SI{ 1.536 \pm0.006}{\milli\meter}$, and $L_{x}^{(2)} = \SI{1.522\pm 0.006}{\milli\meter}$, $L_{y}^{(2)} = \SI{1.525 \pm0.006}{\milli\meter}$.
We studied the dynamics of the lower frequency mode of the two membranes: for the first membrane we measured $\omega_{1} \simeq 2\pi\times\SI{230.795}{\kilo\hertz}$, $\gamma_{1} \simeq 2\pi\times\SI{1.64}{\hertz}$, while for the second $\omega_{2} \simeq 2\pi\times\SI{233.759}{\kilo\hertz}$, $\gamma_{2} \simeq 2\pi\times\SI{9.37}{\hertz}$.
The membrane--cavity is placed in the middle of an optical cavity with empty cavity finesse $\mathcal{F}_0 = \SI{50125\pm25}{}$, which reduces to $\mathcal{F} = \SI{12463\pm13}{}$ when the membrane--sandwich is placed in.
Such finesse corresponds to a cavity intensity decay rate $2\kappa = \tau^{-1} = \mathrm{FSR} / \mathcal{F} \simeq 2\pi\times\SI{134}{\kilo\hertz}$, with $\mathrm{FSR} \simeq 2\pi\times\SI{1.67}{\giga\hertz}$.
%

%
\section{Spectral analysis of the probe beam}
\label{sec:Probe analysis}

Experimentally we have studied the weak driving regime of the onset of synchronization where only one of the two membrane resonators, say first oscillator, enters into a limit cycle through the Hopf bifurcation associated with the parametric instability~\cite{Marquardt2006}. In this regime the other resonator remains in a mixed condition where the modulation of the radiation pressure force induced by the first oscillator is not yet able to prevail over the thermal noise contribution~\cite{Holmes:2012aa,Li2020}.

\begin{figure}[!]
\begin{center}
   {\includegraphics[height=.4\textwidth]{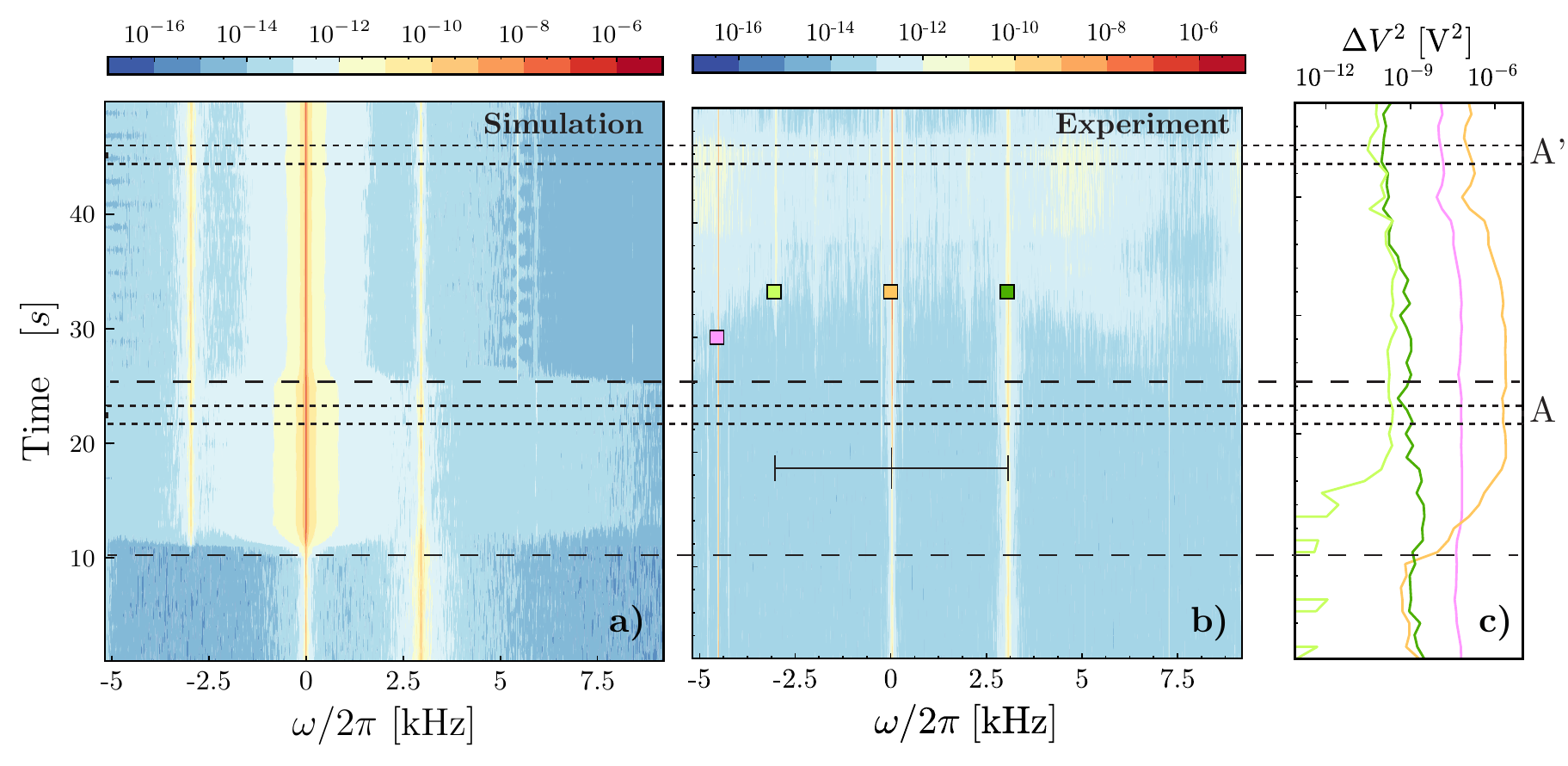}}
 \caption{
  	Spectral noise of the output probe beam centered at the fundamental mode frequency of the first oscillator, which reaches a limit cycle, in a range of time of $\SI{50}{s}$.
	{\bf a)} Numerical simulation of the output field voltage spectral noise, and {\bf b)} experimental voltage spectral noise (VSN in $\si{\square\volt\per\hertz}$).
	After 10 seconds the pump beam is turned on with a power of \SI{4.25}{\micro\watt}, and after 25 seconds increased to \SI{6.0}{\micro\watt}.	
	The optomechanical parameters derived as mean values of the first 8 thermal spectra, are: $\omega_{1} \simeq 2\pi\times\SI{230.795}{\kilo\hertz}$, $g_{1} \simeq 2\pi\times\SI{0.43}{\hertz}$, $\omega_{1} \simeq 2\pi\times\SI{233.759}{\kilo\hertz}$, $g_{2} \simeq 2\pi\times\SI{.70}{\hertz}$.
	{\bf c)} Variances corresponding to the frequency ranges highlighted by the squares symbols:
	orange curve for the first mode;
	dark--green curve for the second mode;
	light--green curve for the sideband at $2\omega_1 - \omega_2$;
	magenta curve for the calibration tone at $\omega_{b} = 2\pi\times\SI{226.000}{\kilo\hertz}$.
	Residual detuning beat tone at $\omega_{det} = 2\pi\times\SI{259.350}{\kilo\hertz}$ is outside the displayed frequency range.
	Areas $A$ and $A^\prime$ indicate the steady--state regime reached for the two power settings.
}
\label{fig:Fig_20200110_Set3_ALL}
\end{center}
\end{figure}

%
In Fig.~\ref{fig:Fig_20200110_Set3_ALL} are compared simulated and experimental results for a set of parameters in this weak driving regime. 
On panel b) is reported the voltage spectral noise (VSN) of the homodyne signal as a function of time, where the frequencies are counted from the frequency of the fundamental mode of the first oscillator, $\omega_{1}$, (marked by an orange square symbol, while the second mode is marked by a dark--green square symbol).
During the first \SI{10}{\second}, the pump beam is turned off, and the VSN shows the thermal displacement of the fundamental modes of the two membranes.
The magenta square symbol marks the external tone used for determining the single--photon optomechanical coupling $g_{1} \simeq 2\pi\times\SI{0.43}{\hertz}$, and $g_{2} \simeq 2\pi\times\SI{0.70}{\hertz}$, and
for calibrating the VSN in displacement spectral noise (DSN)~\cite{Gorodetsky:2010uq}.
Finally, after \SI{10}{\second} the pump beam is turned on at a blue detuning $\Delta_1/2 \pi = \SI{259.350}{\kilo\hertz}$ with a power of \SI{4.25}{\micro\watt} for studying the dynamics of the optomechanical system, and after \SI{25}{\second} increased to \SI{6.0}{\micro\watt}.
Panel c) of Fig.~\ref{fig:Fig_20200110_Set3_ALL} shows the variances corresponding to the integral over the frequency range denoted by the squares symbols in panel b).
Panel a) of Fig.~\ref{fig:Fig_20200110_Set3_ALL} shows the numerical simulation of the non--linear dynamics of the system provided by integration of eqs.~(\ref{eq:c_langevin1})--(\ref{eq:c_langevin}). We note the remarkable agreement between simulated and experimental data.
Results in Fig.~\ref{fig:Fig_20200110_Set3_ALL} shows that when the first membrane reaches a limit cycle, sidebands around its central frequency appears on the output field.
In particular the experimental data and simulations demonstrate the appearance of the sideband due to the second mode (light--green square symbol) at $2\omega_1-\omega_2$.

For a detailed quantitative description of these results, we need to reconsider the non--linear dynamics of the mean cavity--field amplitude $\alpha_2$ of the probe beam described by eq.~(\ref{eq:c_langevin1}), and replace the input noise operator with the input field with a sinusoidal modulation at $\omega_{b}$ for calibration. Moreover, for simplicity we drop everywhere the index $2$ when referring to the cavity probe field, that is $\alpha_2 \to \alpha$, $\kappa_2 \to \kappa$ and so on. We get
\begin{equation}\label{eq:alpha}
	\dot \alpha = -\kappa \alpha
			   + \ii\left[\Delta^{(0)} + 2g_1\text{Re}[\beta_1(t)] + 2g_2\text{Re}[\beta_2(t)]\right]\alpha
			   +E_{in}\,,
\end{equation}
 where $E_{in}  = \sqrt{2\kappa_{in}}\, \text{e}_{in}\exp[-\ii\beta\sin(\omega_{b}t)]$, and $\text{e}_{in} = \sqrt{\calP_{in}/\hbar\omega_L}$. We assume that the behaviour of the two oscillators is also sinusoidal as in eq.~(\ref{ansatz}), i.e., $\beta_j(t) = \beta_{0,j} + |A_j|\exp[\ii(\omega_{mj}t + \phi_j)]$, and that the first oscillator reaches a limit cycle with an amplitude $|A_1|$ for which $g_1|A_1|$ is much larger that  $g_2|A_2|$. The solution of eq.~(\ref{eq:alpha}) can be found considering an expansion in terms of $\epsilon = g_2|A_{2}|/g_1|A_1|$, that is $\alpha = \sum_j \epsilon^j \alpha_j$.
The zero--order solution, $\alpha_0$, satisfies
\begin{eqnarray}\label{eq:alpha_0}
	\dot \alpha_0 = -\kappa\,\alpha_0
			   + \ii\left[\Delta + 2g_1|A_1|\cos(\omega_{1}t + \phi_1)\right]\alpha_0
			   + \sqrt{2\kappa_{in}}\,\text{e}_{in}\ee^{-\ii\beta\sin(\omega_{b}t)}\,,
\end{eqnarray}
where $\Delta = \Delta^{(0)} + 2 g_1\text{Re}[\beta_{0,1}]+  2g_2\text{Re}[\beta_{0,2}]$, and a first--order perturbation solution $\alpha_1$, driven by the amplitude of the second oscillator,
\begin{eqnarray}\label{eq:alpha_1}
	 \dot \alpha_1 = -\kappa\,\alpha_1
			   \!+\! \ii\left[\Delta + 2g_1|A_1|\!\cos(\omega_{1}t \!+\! \phi_1)\right]\alpha_1
			   \!+\! \ii2g_2 |A_2|\cos(\omega_{2}t\! +\! \phi_2)\frac{\alpha_0}{\epsilon}
				\!\,.
\end{eqnarray}
Solution of eq.~(\ref{eq:alpha_0}) provides the leading--order contribution to the cavity response function $\calC = \sum_j \epsilon^j \calC_j$, with  $\calC= \alpha/\text{e}_{in}$,
\begin{eqnarray}\label{eq:C0}
	\calC_0 = \sqrt{2\kappa_{in}}\,
	\sum_{b}J_b(-\beta)\,\ee^{\ii b\,\omega_bt}\,C_b(\xi)
	\,,
\end{eqnarray}
where $\xi = 2g_{1}|A_1|/\omega_{1}$, and
\begin{eqnarray}\label{eq:Cb}
	C_b(\xi) = \sum_{m,n}
	\frac{J_{m-n}(-\xi)J_m(-\xi)}{\ii(m\omega_{1} +b\omega_b) - \calL}\,\ee^{\ii n(\omega_{1}t + \phi_1)},
\end{eqnarray}
with $\calL =i \Delta-\kappa$.
The first--order solution can be found to be
\begin{eqnarray}\label{eq:C1}
	\epsilon\,\calC_1 =\sqrt{2\kappa_{in}}\,
	\sum_{b}J_b(-\beta)
	\,\ee^{\ii b\,\omega_bt}
	\Big[
		&\delta C^{+}_{b}(\xi)
		\,\ee^{\ii(\omega_{2}t + \phi_2)}		 + \delta C^{-}_{b}(\xi)
		 \,\ee^{- \ii(\omega_{2}t + \phi_2)}
	\Big]
	\,,
\end{eqnarray}
where
\begin{eqnarray}\label{eq:Cbpm}
	\delta C^{\pm}_{b}(\xi) = \frac{\ii g_2|A_2|}{\sqrt{2}}\sum_{m,n}
	\frac{J_{m-n}(-\xi)J_m(-\xi)}{\ii(m\omega_{1} + b\omega_b)-\calL}\,\,
	\frac{\,\ee^{\ii n(\omega_{1}t + \phi_1) }}
	{\ii(m\omega_{1} +b\omega_b \pm\omega_{2}) -\calL }
	\,.
\end{eqnarray}
Finally we observe that $\calC$ is composed by a series of sidebands at frequencies $n\omega_{1} + b\omega_{b} \pm \omega_2$, which are reproduced on the reflected output field determined by the input--output relation $\text{e}_{out} = -\text{e}_{in}\exp[-\ii\beta\sin(\omega_{b}t)] + \sqrt{2\kappa_{in}}\alpha$.

The cavity reflection function, $\calR = \text{e}_{out}/\text{e}_{in}  = \sum_j \epsilon^j \calR_j$, has a leading--order contribution
\begin{eqnarray}\label{eq:R0}
	\calR_0
	= \sum_{b}J_b(-\beta)\,\ee^{\ii b\,\omega_bt}\,R_b(\xi)\,.
\end{eqnarray}
We observe that $\calR_0$ is the superposition of the cavity reflection function $R_b(\xi)$ at each sideband frequencies $b\omega_{b}$ of the input field
\begin{eqnarray}\label{eq:Rb}
	R_b(\xi) = -1 + 2\kappa_{in}\,C_b(\xi)\,.
\end{eqnarray}
The first--order contribution is
\begin{eqnarray}\label{eq:R1}
	\calR_1 = \sqrt{2\kappa_{in}}\, \calC_1
	 \,.
\end{eqnarray}
We consider now the modulation amplitude $\beta$, and the amplitude of the second mode, as small perturbations, that is $\beta \ll 1$, and $\calR \simeq \calR_0 + \epsilon\,\calR_1$. In particular we focus on the contributions at DC, and at the five frequencies: $\omega_{1}$, $\omega_{2}$, $\omega_{sm} = 2\omega_{1}-\omega_{2}$, $\omega_{b}$, and $ \omega_{sb} = 2\omega_{1} -\omega_{b}$.

In this case we have
\begin{eqnarray}\label{eq:calH}
	\calR \simeq
		\calR_{DC}
	  &&+\calR_+(\omega_{1}) \,\ee^{\ii(\omega_{1}t + \phi_1)} + \calR_-(\omega_{1})\,\ee^{-\ii(\omega_{1}t + \phi_1)}
		\nonumber \\
	  &&+\calR_+(\omega_{2}) \,\ee^{\ii(\omega_{2}t + \phi_2)} + \calR_-(\omega_{2})\,\ee^{-\ii(\omega_{2}t + \phi_2)}
		\nonumber \\
	  &&+\calR_+(\omega_{sm}) \,\ee^{\ii(\omega_{sm}t + \phi_{sm})} + \calR_-(\omega_{sm})\,\ee^{-\ii(\omega_{sm}t + \phi_{sm})}
		\nonumber \\
	  &&+\calR_+(\omega_{b}) \,\ee^{\ii\omega_{b}t} + \calR_-(\omega_{sb})\,\ee^{-\ii\omega_{b}t}
		\nonumber \\
	  &&+\calR_+(\omega_{sb}) \,\ee^{\ii(\omega_{sb}t + \phi_{sb})} + \calR_-(\omega_{sb})\,\ee^{-\ii(\omega_{sb}t + \phi_{sb})}
	\,,
\end{eqnarray}
The contribution at DC is provided by eq.~(\ref{eq:R0}) for $b = 0$, and the time--independent term ($n = 0$) of eq.~(\ref{eq:Cb})
\begin{eqnarray}\label{eq:ImRDC}
	\calR_{DC}  &=& J_0(-\beta)
	\left[
	-1 + 2\kappa_{in}\sum_{m}\frac{J^2_{m}(-\xi)}{ \ii m\omega_{1}- \calL}
	\right]
	\,;
\end{eqnarray}
at  $\omega_{1}$ is given by eq.~(\ref{eq:R0}) with $b = 0$, and in eq.~(\ref{eq:Cb}) the term with $n = \pm 1$
\begin{eqnarray}\label{eq:ImRm1}
	\calR_\pm(\omega_{1})  &=& J_0(-\beta)\,
		2\kappa_{in}\sum_{m}\frac{J_{m}(-\xi)J_{m\mp1}(-\xi)}{\ii m\omega_{1} - \calL}
			\,;
\end{eqnarray}
at  $\omega_{2}$ is provided by eq.~(\ref{eq:C1}) with $b = 0$, and in eq.~(\ref{eq:Cbpm}) the term with $n = 0$
\begin{eqnarray}\label{eq:ImRm2}
	\calR_\pm(\omega_{2})  = J_0(-\beta)\,2\kappa_{in}\,
		\frac{\ii g_2|A_2|}{\sqrt{2}}\sum_{m}
	\frac{J^2_{m}(-\xi)}{\ii m\omega_{1} - \calL}\,
	\frac{1}{\ii (m\omega_{1} \pm \omega_{2})- \calL}
	\,;
\end{eqnarray}
for  $ \omega_{sm} = 2\omega_{1} -\omega_{2}$,  $b = 0$ in eq.~(\ref{eq:C1}), and $n = \mp 2$ in eq.~(\ref{eq:Cbpm})
\begin{eqnarray}\label{eq:ImRsb}
	\calR_\pm(\omega_{sm}) \!=\! J_0(-\beta)\, 2\kappa_{in}
		\frac{\ii g_2|A_2|}{\sqrt{2}}\sum_{m}&&
	\frac{J_{m}(-\xi)J_{m\mp2}(-\xi)}{ \ii m\omega_{1} - \calL}
				\frac{1}{ \ii(m\omega_{1} \mp \omega_{2})-\calL}
	\,;
\end{eqnarray}
at  $\omega_{b}$ is provided by eq.~(\ref{eq:R0}) with $b = \pm 1$, and in eq.~(\ref{eq:Cb}) the term with $n = 0$
\begin{eqnarray}\label{eq:ImRb}
	\calR_\pm(\omega_b)  &=& J_{\pm 1}(-\beta)
		\left[
		-1 + 2\kappa_{in}
		\sum_{m}\frac{J^2_{m}(-\xi)}{\ii (m\omega_{1} \pm \omega_b) - \calL}
		\right]
	\,;
\end{eqnarray}
and the sideband at $ \omega_{sb} = 2\omega_{1} -\omega_{b}$ for $b = \pm 1$, and $n = 0$
\begin{eqnarray}\label{eq:ImRbb}
	\calR_\pm(\omega_{sb}) = J_{\mp 1}(-\beta)\, 2\kappa_{in}\,
		\sum_{m}&&
	\frac{J_{m}(-\xi)J_{m\mp2}(-\xi)}{ \ii(m\omega_{1} \mp \omega_{b}) - \calL}
	\,.
\end{eqnarray}

\begin{figure}[t!]
\begin{center}
   {\includegraphics[height=.29\textwidth]{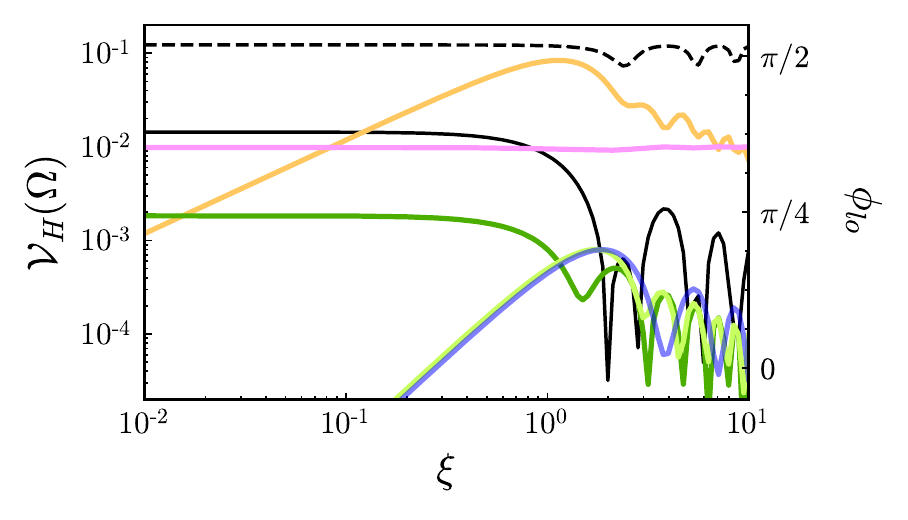}}
 \caption{Normalized amplitude spectral noise, $\calV_H(\Omega)$, obtained by using the expression in eq.~(\ref{eq:VHOmega}), and the experimental parameters measured for the results reported in Fig.~\ref{fig:Fig_20200110_Set3_ALL}, considering a small nonzero probe detuning $\Delta \sim 2\times \SI{3.9}{\kilo\hertz}$, $\beta = \SI{2e-2}{\radian}$, and the dimensionless amplitude $A_{2} = \sqrt{k_BT/m\omega_{2}^2}/2\, x_{\rm zpf}\sim 3634$.
 	Black curve is the DC contribution~(\ref{eq:ImRDC});
	orange curve at $\omega_{1}$, eq.~(\ref{eq:ImRm1});
	dark--green curve at $\omega_{2}$, eq.~(\ref{eq:ImRm2});
	light--green curve at $\omega_{sm}  = 2\omega_{1} - \omega_{2}$, eq.~(\ref{eq:ImRsb});
	magenta curve at $\omega_{b}$, eq.~(\ref{eq:ImRb});
	blue line curve at $\omega_{sb}  = 2\omega_{1} - \omega_{b}$, eq.~(\ref{eq:ImRbb}).
	Black dashed curve is the local oscillator phase, $\phi_{lo}$, which, for our experimental parameters, can be considered $\simeq \pi/2$.
}
\label{fig:S_NEW}
\end{center}
\end{figure}

The homodyne technique is implemented by mixing on a beam--splitter the reflected field $\text{e}_{out}$ with an intense local oscillator $\text{e}_{lo}\,\ee^{\ii\phi_{lo}}$, and detecting the fields at the output of the beam--splitter, i.e.,  $\text{e}_j = [\text{e}_{lo}\,\ee^{\ii\phi_{lo}} +(-1)^j \text{e}_{out}]/\sqrt{2}$. The output power are $\calP_{\pm} = |\text{e}_{lo}\,\ee^{\ii\phi_{lo}} \pm \text{e}_{out}|^2/2$, and the differential current
\begin{eqnarray}
	I_D 	= S\,[\mathcal{P}_{+}- \mathcal{P}_{-}]
		= 2S\,\sqrt{\mathcal{P}_{lo}\mathcal{P}_{in}}\,\text{Re}\left[\calR\,\ee^{-\ii\phi_{lo}} \right]\,,
\end{eqnarray}
where $S$ is the sensitivity of the photodiodes, and the phase $\phi_{lo}$ represents the controllable phase difference between the local oscillators and the reflected field.
The local oscillator phase is locked to have zero DC signal, which turns out to be
$\phi_{lo} = \arctan\big[\text{Re}(\calR_{DC} )/\text{Im}(\calR_{DC})\big]$. This phase is plotted in Fig.~\ref{fig:S_NEW} for our experimental parameters, showing that it does not deviate from the optimal value of $\pi/2$.
The error signal, assuming optimal detection for the local oscillator phase $\phi_{lo} \simeq \pi/2$, is
\begin{eqnarray}
	V_H(t) = g_TS\,2\sqrt{\mathcal{P}_{lo}\mathcal{P}_{in}}\,\,\text{Im}\left[\calR \right]\,,
\end{eqnarray}
where $g_T$ is the transimpedance gain.
This voltage contains the signature of any modulation frequency of the reflected field, provided that it falls within the bandwidth of the electronic system.
The single--sided  power spectral density  $\calS_{W}(\omega)  =\int d\tau \ee^{\ii\omega \tau}\langle V_H(t+\tau)V_H(t)\rangle_t/R_0$ on a termination resistor $R_0$, provides a normalized amplitude spectral noise for each well separated frequency, that is
\begin{eqnarray}\label{eq:VHOmega}
	 \calV_H(\Omega) = \sqrt{\frac{\calS_{W}(\Omega)}{\calS_0} } =
	 \frac{1}{2}\,\left | \calR_+(\Omega) - \calR_-^\ast(\Omega) \right |
	\,,
\end{eqnarray}
with $ \calS_0 = (g_T\,2S)^2\,\mathcal{P}_{lo}\mathcal{P}_{in}/R_0$.
In Fig.~\ref{fig:S_NEW} are reported the amplitude spectral noise at the frequencies of interest for our experiment.

We notice that for probe detuning $\Delta = 0$, and when the cavity field is weakly modulated at frequency $\omega_{1}$, that is $\xi \ll 1$, the output signal is linear in the displacement
\begin{eqnarray}
	\calR_+(\omega_{1}) = - [\calR_-(\omega_{1})]^\ast  \sim
			-\frac{\xi}{2}
			\frac{2\kappa_{in}}{\kappa}\frac{\ii\omega_{1}}{\kappa + \ii\omega_1}\,,
				\hspace{1.5cm}						
\end{eqnarray}
and the amplitude spectral noise is
\begin{eqnarray}
	 \sqrt{\calS_{W}(\omega_1) }\Big[\si{\watt\per\sqrt{\hertz}}\Big]	
		= \frac{g_TS}{\sqrt{R_0}}
		  \cdot 2\sqrt{\mathcal{P}_{lo}\mathcal{P}_{in}}\cdot\frac{\mathcal{F}}{\lambda_0}
			\frac{\delta \tilde x(\omega_1)\Big[\si{\meter\per\sqrt{\hertz}}\Big]}{\sqrt{1 + \omega_1^2/\kappa^2}}
	 		\!\cdot\!\eta	\,\,\,,
\end{eqnarray}
where $\calF = {\rm FSR}/2\kappa$, $\xi = \delta \tilde x(\omega_1)g_1/\omega_1x_{\rm zpf}$, and $\eta = [2\kappa_{in}/\kappa]\cdot[g_1\,\lambda_0/2\,{\rm FSR}\,x_{\rm zpf}]\sim 0.25\cdot0.24\sim0.06$ for our setup.
The average power falling on each photodiode is approximately $\mathcal{P}_{lo}/2$. The shot noise in the differential signal has a flat spectrum with spectral density  $S^{\rm sn}_{\mathcal{P}\mathcal{P}}\sim 2\times2\hbar\omega_L\, \mathcal{P}_{lo}/2$, which sets a limit to the sensitivity of the detection~\cite{Black:2001kx,Schliesser:2008hc}
\begin{eqnarray}
	\delta \tilde x(\omega_1)\Big[\si{\meter\per\sqrt{\hertz}}\Big]
		=  \frac{1}{\sqrt{2\mathcal{P}_{in}/\hbar\omega_L}}\,\frac{\lambda_0}{\mathcal{F}}\frac{1}{\eta}
			 	   \sqrt{1 + \frac{\omega_1^2}{\kappa^2}}
		\sim \SI{4e-16}{\meter\per\sqrt{\hertz}}
	 			\,,
\end{eqnarray}
and reproduces the shot--noise limited displacement detection filtered by the cavity response, for which the shot--noise limited sensitivity is not flat in the spectrum (Mizuno's sensitivity theorem~\cite{Mizuno:1995aa}). The cavity response length in this linear detection regime might be estimated as $\lambda_0/2\calF \simeq \SI{43}{\pico\meter}$, which corresponds to $\xi_{cav}=2g_1/\omega_1 \cdot \lambda_0/2\calF x_{\rm zpf} \simeq \num{0.284}$.

On the contrary, when the excited mechanical mode reaches a limit cycle with large amplitude, that is for $\xi$ approaching unity, the reflected signal, due to the non--linear response of the cavity, presents a reduction of the signal of the unexcited mode and the appearance of a sideband at $\omega_{sm}$ and $\omega_{sb}$ (see Fig.~\ref{fig:S_NEW}).
In this case any attempt to determine the mechanical displacement from the measured phase of the output field requires careful attention, also because the calibration tone, which is implemented by modulation of the input field, is essentially unaffected (see magenta line in Fig.~\ref{fig:S_NEW}).
\begin{figure}[t!]
\begin{center}
   {\includegraphics[height=.4\textwidth]{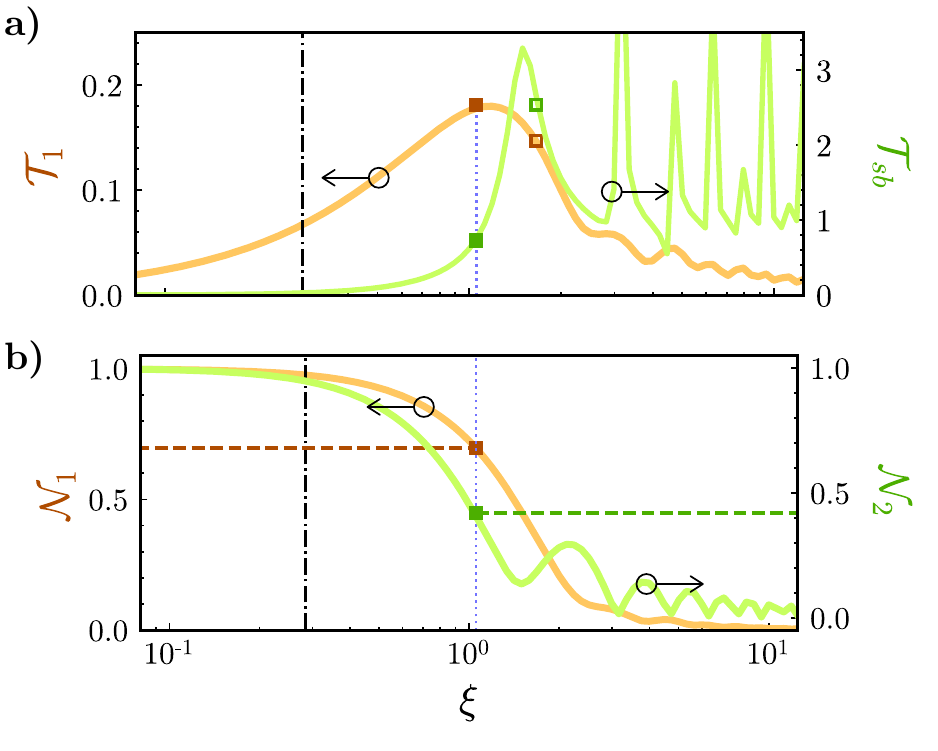}}
 \caption{
	 {\bf a)} Ratios $\calT_1$ of eq.~(\protect\ref{ratiosT}), orange curve, and $\calT_2$, light--green curve, as a function of $\xi$.  	
	Filled square symbols correspond to the data indicated by the areas $A$ in Fig.~\protect\ref{fig:Fig_20200110_Set3_ALL} for the measurements in frequency domain, and the one reported in Fig.~\ref{fig:Fig_20200110_Set3_Brownian_2D_Volt} for the measurements in time domain, with an estimated $\xi_1^{\rm st} \simeq 1.05$, which corresponds to $q_1^{\rm st} = \xi_1^{\rm st}\,\omega_1\,x_{\rm zpf}/g_1 \simeq \SI{263}{\pico\meter}$.
	Open square symbols correspond to the data indicated in Fig.~\ref{fig:Fig_20200110_Set3_ALL} by the areas $A^\prime$, corresponding to $\xi_{A^\prime} \simeq \num{1.66}$.
	{\bf b)} Ratios $\calT_1\xi^{-1}$, and $\calT_2$ normalized to their maximum values, and indicated as $\calN_1$, orange curve, and $\calN_2$, light--green curve, respectively.
 	For $\xi_1^{\rm st} \simeq 1.05$ estimated in panel a), and evidenced by the blue dotted line, the attenuations due to the optical readout  (as explained in the text) are $\calN_1  \simeq  0.70$ and $\calN_2 \simeq 0.42$, for which the observable displacement is $q^{\rm ob}_1 = q^{\rm st}_1\,\calN_1 \simeq \SI{183}{\pico\meter}$.
	Dot--dashed line corresponds to the cavity response length, that is the maximum displacement detectable in the linear regime, $\xi_{cav}\simeq \num{0.284}$, as explained in the text.
}
\label{fig:Ratio_ALL}
\end{center}
\end{figure}
In general, two main ratios of the normalized spectral amplitudes $\calV_H(\Omega)$, which are independent from $\beta$, $\kappa_{in}$, and from the coupling term with the second mechanical mode, $g_2|A_2|$, describe the non--linear dynamics of the optomechanical system in terms of the normalized mechanical amplitude of the first mechanical oscillator $\xi$: i) the ratio derived by eq.~(\ref{eq:ImRm1}) and the product of $\beta$ and eq.~(\ref{eq:ImRb}); ii) the ratio derived by eq.~(\ref{eq:ImRm2}) and eq.~(\ref{eq:ImRsb})
\begin{eqnarray}
	\calT_1 = \frac{\calV_H(\omega_{1})\cdot\beta}{\calV_H(\omega_b)},
		\hspace{2cm}
	\calT_{sb} = \frac{\calV_H(\omega_{sb})}{\calV_H(\omega_{2})}
	\label{ratiosT}	 			\,.
\end{eqnarray}
In panel a) of Fig.~\ref{fig:Ratio_ALL} are reported these ratios that allow us to estimate $\xi$ for our experimental realisations.
In panel b) the ratios $\calT_1(\xi)\xi^{-1}$, and $\calT_2(\xi) = \calV_H(\omega_{2})\cdot\beta/\calV_H(\omega_b)$ (note that $\calT_2$ does depend on $g_2|A_2|$), normalized to their maximum values, which are reached for $\xi \rightarrow 0$, are shown as $\calN_1$, and $\calN_2$, respectively.
By definition, they represent the correction factors that relate the displacement amplitudes detected via the reflected probe spectrum, to the effective displacements amplitudes. $\calN_1$, and $\calN_2$ are equal to $1$, corresponding to the usual linear detection regime, for $\xi \ll \num{1}$.
In the present case instead, for the results reported in Fig.~\ref{fig:Fig_20200110_Set3_ALL}, and Fig.~\ref{fig:Fig_20200110_Set3_Brownian_2D_Volt} we estimate $\xi_1^{\rm st} \simeq  1.05$, $\calN_1\simeq  0.70$ and $\calN_2 \simeq 0.42$. Such analysis allows us to deduce the effective limit cycle displacement amplitude of the first oscillator to be $q_1^{\rm st} = \xi_1^{\rm st}\,\omega_1\,x_{\rm zpf}/g_1 \simeq \SI{263}{\pico\meter}$, and the observed limit cycle displacement amplitude $q^{\rm ob}_1 = q_1^{\rm st}\,\calN_1 \simeq \SI{183}{\pico\meter}$. This analysis does not allow us, for the moment, to draw any quantitative conclusion for what concerns the second oscillator. We will be able to do that in the following Section, when we will analyse the experimental time traces. We anticipate here that also the optical detection of the unexcited oscillator is affected by the nonlinear response of the cavity, and the complementary analysis in time domain will allow us to determine the proper correction to the calibration readout ${\cal N}_2$. 

\section{Non--linear mechanical  dynamics at the onset of synchronization}
\label{nonlinear}
In contrast to the systems implemented in~\cite{Sheng:2020aa,Kemiktarak:2014cq}, our
high finesse optical system, used for detection, does not allow us to reveal the effective motion of each membrane independently.
However, by using the above analysis of the reflected spectrum, and the numerical simulations, it is possible to unambiguously infer from the homodyne detection of the probe beam shown in Fig.~\ref{fig:Fig_20200110_Set3_ALL} that the dynamics of the two membranes is characterized in this parameter region by a pre--synchronisation regime.
In fact the numerical integration of eqs.~(\ref{eq:c_langevin1})--(\ref{eq:c_langevin}) with the parameters reported in Table~\ref{tb:tab1}, which refers to the results of Fig.~\ref{fig:Fig_20200110_Set3_ALL}, also shows that when the output probe beam exhibits sidebands around the frequency of the fundamental mode of the first oscillator [see Fig.~\ref{fig:Fig_20200110_Set3_ALL}a)], the second oscillator has a nonzero amplitude of oscillation at the frequency of the first oscillator, that is, it starts to synchronize with the first mode.
\begin{figure}[t!]
\begin{center}
   {\includegraphics[height=.4\textwidth]{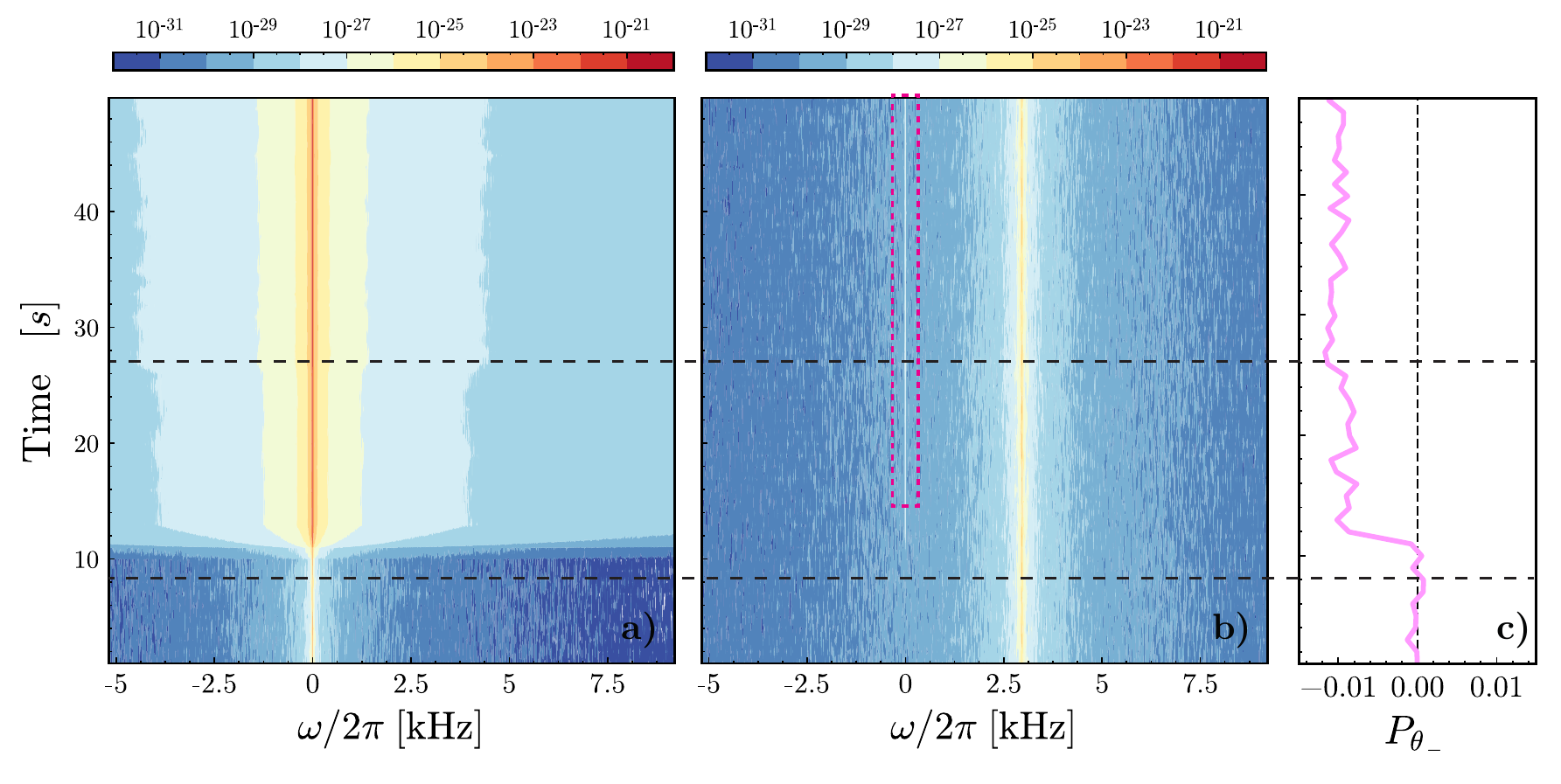}}
 \caption{
  	{\bf a)}  Numerical simulation of the DSN of the fundamental mechanical mode of the first oscillator for the experimental parameters of Fig.~\ref{fig:Fig_20200110_Set3_ALL}. 
  	After $\SI{10}{\second}$ the pump beam is turned on, and after $\SI{25}{\second}$ increased.
	{\bf b)}  Numerical simulation of the DSN of the fundamental mechanical mode of the second oscillator. While the first oscillator reaches a limit cycle, the second oscillator starts to present an amplitude at the frequency of the first one, highlighted by the magenta box.
	{\bf c)} Synchronization measure as defined in eq.~(\ref{eq:Ptheta}). Phase anti--correlation between the two oscillators increases with the pump beam power.
	}
\label{fig:Fig_20200110_Set3_q1_q2}
\end{center}
\end{figure}
This is emphasised by the magenta box in panel b) of Fig.~\ref{fig:Fig_20200110_Set3_q1_q2}, which shows the numerical simulations of the spectra of the fundamental mechanical mode of the two oscillators for the experimental parameters of Table~\ref{tb:tab1}. A more quantitative description of such pre-synchronization process of the second resonator with the excited first one, acting as the ``master'' oscillator, is provided by the synchronization measure~\cite{Li2020}
\begin{equation}
	P_{\theta_-}(t)=\frac{1}{\Delta t} \int^{t+\Delta t/2}_{t-\Delta t/2}\cos\left[\theta_1(t)-\theta_2(t)\right]dt.
\label{eq:Ptheta}
\end{equation}
where $\theta_j(t)=\arg[\beta_j(t)]$, reported in Fig.~\ref{fig:Fig_20200110_Set3_q1_q2}c), which shows an increase of the phase anti--correlation between the two oscillators. The effect is small due to the weak driving regime, but it is nonetheless unambiguously present.

\begin{table}[b!]
\caption{Optomechanical parameters for the results reported in Fig.~\ref{fig:Exp_Theory_ALL}.\vspace{.25cm}}
        \centering
       \begin{tabular}{cccc}
         \hline
         $\omega_{1}$  & & $2\pi\cdot\SI{230.795}{\kilo\hertz}$ \\
         $\omega_{2}$ & & $2\pi\cdot\SI{233.759}{\kilo\hertz}$ \\
         $\omega_{b}$ & & $2\pi\cdot\SI{225.350}{\kilo\hertz}$ \\
         $\Delta_1$ & & $2\pi\cdot\SI{259.350}{\kilo\hertz}$ \\
         $\Delta_2$ & & $2\pi\cdot\SI{3.9}{\kilo\hertz}$ \\
         $g_{1}$ & & $2\pi\cdot\SI{0.4225}{\hertz}$  \\
         $g_{2} $ & & $2\pi\cdot\SI{0.6965}{\hertz}$   \\
         $\gamma_{1} $ & & $2\pi\cdot\SI{1.64}{\hertz}$ \\
         $\gamma_2 $ & & $2\pi\cdot\SI{9.37}{\hertz}$ \\
         $\kappa_{loss}$ & & $2\pi\cdot\SI{50.35}{\kilo\hertz}$ \\
         $\kappa_{in}$ & & $2\pi\cdot\SI{8.35}{\kilo\hertz}$ \\
         $\kappa_{ex}$ & &  $2\pi\cdot\SI{58.7}{\kilo\hertz}$ \\
         ${\rm FSR}$ & &  $2\pi\cdot\SI{1.67}{\giga\hertz}$ \\
         $P_{pump}$ & & $\SI{4.25}{\micro\watt}$ \\
         $P_{probe}$ & & $\SI{5.9}{\micro\watt}$ \\
         $\lambda_0$ & & $\SI{1064}{\nano\meter}$ \\
         \hline
       \end{tabular}
       \label{tb:tab1}
\end{table}

We are also able to provide a consistent analytical description of this pre-synchronization dynamics of the two membrane modes starting from the slowly varying amplitude equations~(\ref{eq:first-order finnal0})-(\ref{eq:first-order finnal1}). To study the regime when the first oscillator reaches a limit cycle while the second is not excited, it is convenient to take $\omega_{\rm ref}=\omega_{1}$ as a reference in eqs.~(\ref{eq:first-order finnal0})-(\ref{eq:first-order finnal1}), so that $\Delta \omega_1 = 0$ and $\Delta \omega_2 = \Delta \omega$. One can make quantitative predictions on such a regime assuming that $|A_1| \gg |A_2|, \sqrt{2 \bar{n}_1}$. Moreover, in our experiment the optical noise is negligible and we will not consider the terms associated with $\eta_i^{opt}(t)$.
With the above approximations, one can neglect both thermal noise and $A_2$ contributions in Eq.~(\ref{eq:first-order finnal0}), which becomes
\begin{eqnarray}\label{eqampliappr1}
  \dot{A}_1(t)	\!=\!\Big[-\gamma_1+\ii d_1(|A_1|)\Big] A_1(t)
			\!=\! -\Big[\gamma_1^{eff}(|A_1|) - \ii \Delta\omega_1^{eff}(|A_1|)\Big] A_1(t),
\end{eqnarray}
where we have made explicit the dependence of $d_1$ on $|A_1|$, and defined $\Delta\omega_1^{eff}(|A_1|) = -\text{Re}[d_1(|A_1|)]$, and $\gamma_1^{eff}(|A_1|) = \gamma_1 + \text{Im}[d_1(|A_1|)]$. The effective mechanical damping can be cast as
\begin{eqnarray}
	\gamma_1^{eff}(|A_1|) = \gamma_{1}
	\left[
		1 +  \frac{g_1}{\gamma_1|A_1|}\text{Im}\left[E_1^2\,\Sigma_1 + E_2^2\,\Sigma_2\right]
	\right]	\,,
\label{eq:first-order finnal}
\end{eqnarray}
where we have used the fact that in the considered regime $|A_i^b| \simeq |A_1| (g_{i1}/g_i^b)$, $\xi_j \simeq \xi_1 = 2g_1|A_1|/\omega_1$, assumed $g_{i1} \simeq g_1$, and
\begin{eqnarray}
	\Sigma_j \equiv \Sigma(\xi_1,\kappa_i,\Delta_i) = \sum_n\frac{J_n\left(-\xi_1\right)J_{n+1}\left(-\xi_1\right)}
					{[\ii n{\omega_{1}}-\calL_j][-\ii (n+1)\omega_{1}-\calL_j^*]}.
\end{eqnarray}\label{eq:Sigmaj}
We note that such approximation implies a regime where the second mode is still dominated by a thermal dynamics, i.e., pre--synchronized regime; on the contrary, if also the second mode would have reached a limit cycle, synchronized with the first, the amplitude $g_2A_2$ would have not been negligible anymore with respect to $g_1A_1$, and the dynamics would have been governed by the more general eqs.~(\ref{eq:first-order finnal0})--(\ref{eq:first-order finnal1})~\cite{Sheng:2020aa,Kemiktarak:2014cq}.
Eq.~(\ref{eqampliappr1}) can be solved by rewriting it in terms of modulus and phase, $A_1 = I_1 \ee^{\ii \phi_1}$,
\begin{eqnarray}\label{eqamplimodu}
  \dot{I}_1(t)&=&-\gamma_1^{eff}(I_1)\, I_1(t),\\
  \dot{\phi}_1(t)&=& \Delta\omega_1^{eff}(I_1)\,. \label{eqamplifase}
\end{eqnarray}
After a transient these equations yield a steady state with a constant radius of the limit cycle of the first oscillator, $I_1^{\rm st}$, corresponding in our case of not too strong driving, to the smallest positive root of the implicit equation $\gamma_1^{eff}(I_1^{\rm st})=0$, which can be cast as
\begin{eqnarray}
	  a\,|\xi_1^{\rm st}| =  - \text{Im}\left[E_1^2\,\Sigma(\xi_1^{\rm st},\kappa_1,\Delta_1)
	  			 + E_2^2\,\Sigma(\xi_1^{\rm st},\kappa_2,\Delta_2)\right] \,,
\label{eq:first-order finnal2}
\end{eqnarray}
with $\xi_1^{\rm st} = 2g_1I_1^{\rm st}/\omega_1$, and $a = \omega_1\gamma_{1}/2g_1^2$.
\begin{figure}[t!]
\begin{center}
   {\includegraphics[height=.3\textwidth]{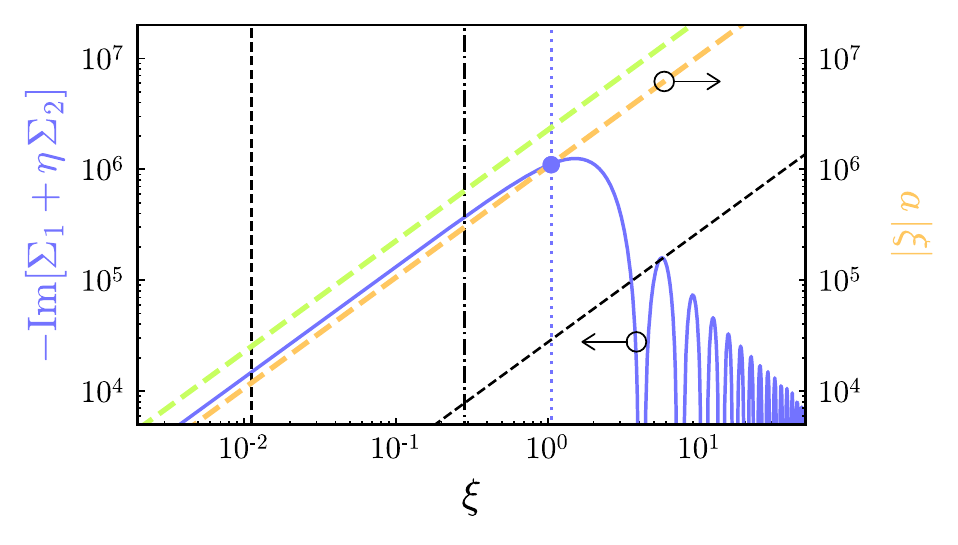}}
 \caption{
 	Steady--state solution for the mechanical displacement amplitude that reaches a limit cycle. The intersection of right and left side of eq.~(\ref{eq:first-order finnal2}) for the parameters of Table~\ref{tb:tab1}, which are reported as solid blue, and dashed orange curves, respectively, determines the steady--state value of $\xi_1^{\rm st}$.
	We determine $\xi_1^{\rm st} = \num{1.054}$, and an effective steady--state amplitude $q^{\rm st}_1 = 2|A_1|x_{\rm zpf}  = \SI{263.0}{\pico\meter}$, which confirms the results shown in Fig.~\ref{fig:Ratio_ALL}.
	The vertical black dashed, and dot-dashed lines represent the values for the thermal displacement  $q_{th} = \sqrt{k_B T/m_1\omega_1^2} \simeq \SI{3.365}{\pico\meter}$ corresponding to $\xi_{th} \simeq \num{0.0112}$, and for the cavity response length
	$\lambda_0/2\calF \simeq \SI{43}{\pico\meter}$, corresponding to $\xi_{cav} \simeq \num{0.284}$, respectively.
	The oblique light--green line represents the left term in eq.~(\ref{eq:first-order finnal2}) for the second less coupled mode, for which the equation is not satisfied for the parameters in Table~\ref{tb:tab1}. In fact, the threshold power, that is the minimum power for finding a root, equivalently for the optical damping to exceed the intrinsic one, for the first mode is $\sim \SI{3}{\micro\watt}$, while for the second $\SI{6.75}{\micro\watt}$. For power larger than $\SI{6.75}{\micro\watt}$ both modes might establish a limit cycle~\cite{Kemiktarak:2014cq}.
	The oblique black dashed line indicates the boundary between the region with only one solution and with multiple solutions, that is the multistability parameter region, which occurs for a pump power larger than $\sim\SI{667}{\micro\watt}$.
}
\label{fig:Sigma_ALL}
\end{center}
\end{figure}
As a consequence, at long times, $\phi_1^{\rm st}(t) \simeq t\Delta\omega_{1}^{\rm st}$ with $ \Delta\omega_{1}^{\rm st} = \Delta\omega_1^{eff}(I_1^{\rm st})$ so that $A_1(t)  \simeq   A_1^{\rm st}(t)= I_1^{\rm st} \exp[{\ii t\Delta\omega_{1}^{\rm st}] }$.

\begin{figure}[b!]
\begin{center}
   {\includegraphics[height=.475\textwidth]{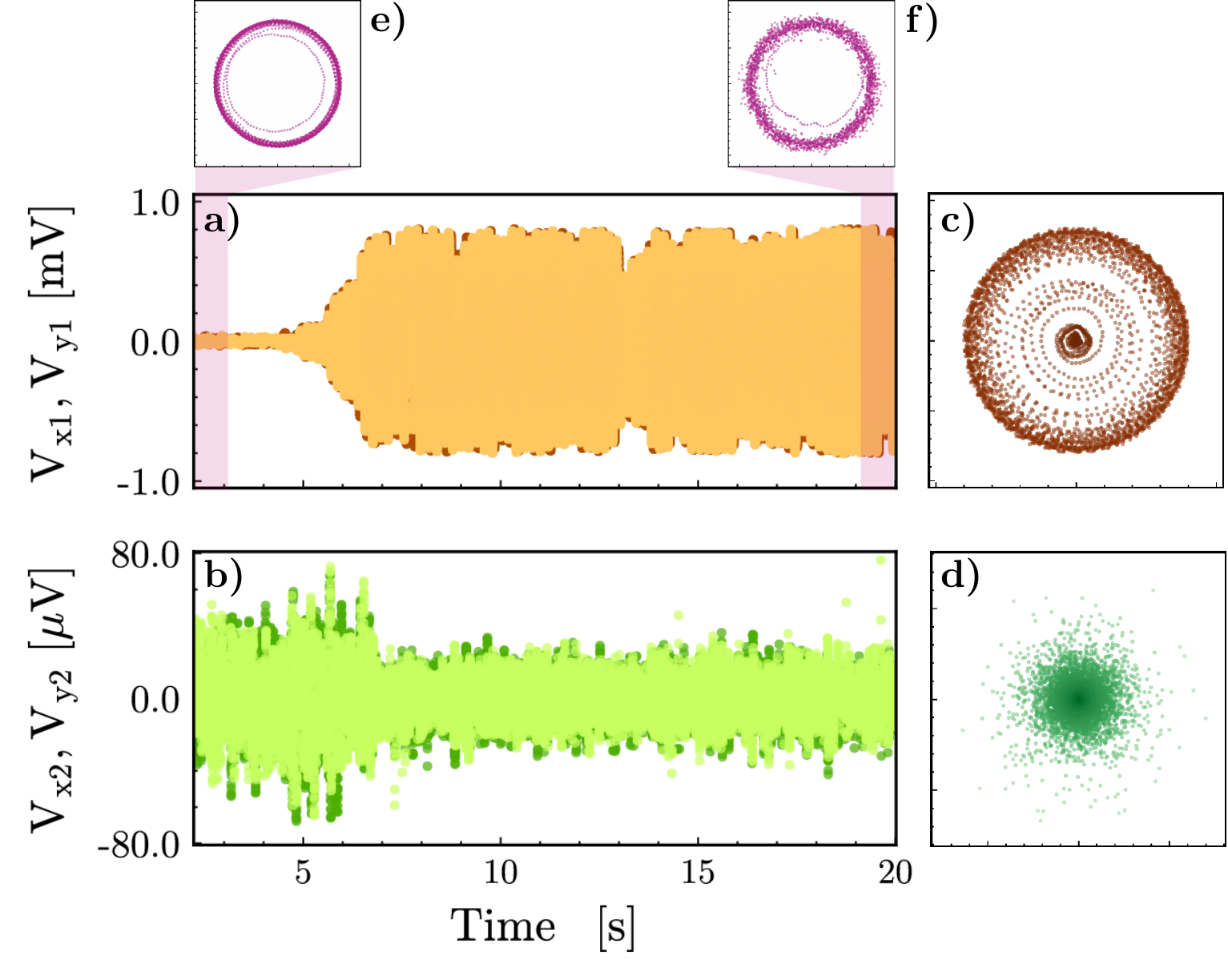}}
 \caption{
	{\bf a)} and {\bf b)} Voltage quadratures, ${\rm V}_{{\rm x}j}$ and ${\rm V}_{{\rm y}j}$, as a function of time for the fundamental modes of the two oscillators. {\bf c)} and {\bf d)} Phase--space distributions associated with the voltage quadratures. {\bf e)} and {\bf f)} Phase--space distribution of the calibration tone before and after the pump is turned on. Optomechanical parameters are the same as in Table.~\ref{tb:tab1}.
}
\label{fig:Fig_20200110_Set3_Brownian_2D_Volt}
\end{center}
\end{figure}

In Fig.~\ref{fig:Sigma_ALL} we show the left and right side of eq.~(\ref{eq:first-order finnal2}) for the experimental parameters of Table.~\ref{tb:tab1}, which provides the optomechanical parameters for the results reported in Fig.~\ref{fig:Fig_20200110_Set3_ALL}, and Fig.~\ref{fig:Fig_20200110_Set3_Brownian_2D_Volt}.
We infer from the intersection point, which corresponds to find the smallest root of $\gamma_1^{eff}(I_1^{\rm st})=0$, a value $\xi_1^{\rm st} = \num{1.054}$, a steady--state displacement amplitude $q^{\rm st}_1 = 2|A_1|x_{\rm zpf} = \SI{263.0}{\pico\meter}$, and $\Delta\omega_1^{eff} = -2\pi\cdot\SI{.04}{\hertz}$, confirming the value of $q_1^{\rm st}$ deduced in Fig.~\ref{fig:Ratio_ALL}.
We emphasise once more that, as shown in Fig.~\ref{fig:Ratio_ALL}, for $\xi \ll 1$ the spectral amplitude of the sideband of the output field is linear with $\xi$ and provides a direct measurement of the mechanical position coordinate $q_1$; on the contrary for $\xi\geq 1$ linearity is no more valid and a proper correction factor should be considered. In our case, as obtained in Fig.~\ref{fig:Ratio_ALL}, the theoretical correction factor is $\calN_1 \simeq \num{0.70}$, corresponding to an expected observable stationary limit cycle amplitude of $q^{\rm ob}_1 \simeq \SI{183}{\pico\meter}$ fully consistent with the analysis described in the previous section. It is worth noting that, due to the oscillating behaviour of the Bessel functions, eq.~(\ref{eq:first-order finnal2}) may have more than one solution at sufficiently large power (see below the oblique black dashed line in Fig.~\ref{fig:Sigma_ALL}), corresponding to the multistability phenomenon analysed in Ref.~\cite{Marquardt2006} and experimentally verified in Ref.~\cite{Krause:2015aa}.

We now consider the dynamics of the second oscillator inserting the steady--state solution for $A_1(t)$ into Eq.~(\ref{eq:first-order finnal1}), which becomes
\begin{eqnarray}\label{eqampliappr2}
	\dot{A}_2(t)=\left[-\gamma_2-\ii \Delta \omega +\ii d_2(I_1^{\rm st})\right] A_2(t) +\ii d_{12}(I_1^{\rm st}) A_1^{\rm st}(t) +\sqrt{2\gamma_2}\beta^{in}_2(t).
\end{eqnarray}
The stationary solution can be obtained via Fourier transform and it can be written as
\begin{equation}\label{a2stat}
	A_2(t) =\frac{\ii d_{12}(I_1^{\rm st})}{\gamma_2^{eff}+\ii \Delta\bar\omega_2^{eff}}A_1^{\rm st}(t)+\sqrt{2\gamma_2}\int_0^t ds\, \ee^{-\left(\gamma_2^{eff}+\ii \Delta \omega_2^{eff}\right)s} \beta^{in}_2(t-s),
\end{equation}
where $\gamma_2^{eff}=\gamma_2+\text{Im}[d_2(I_1^{\rm st})]$ is positive, i.e., the second resonator is still damped despite the pump driving and it is not driven into a limit cycle, $\Delta\bar\omega_2^{eff} =\Delta\omega + \text{Re}[d_1(I_1^{\rm st})-d_2(I_1^{\rm st})]$, and $\Delta \omega_2^{eff} =\Delta\omega - \text{Re}[d_2(I_1^{\rm st})]$. Therefore the first term on the right hand side of Eq.~(\ref{a2stat}) is the synchronized component oscillating at the same frequency of the first master oscillator [see Fig.~\ref{fig:Fig_20200110_Set3_q1_q2}b)], while the second term is the thermal noise component at its natural frequency. This equation describes how the second resonator is driven towards synchronization with the first resonator, and full synchronization and phase locking is achieved when the thermal contribution is negligible, i.e., when $ |d_{12}(I_1^{\rm st})|^2 I_1^{\rm st,2} \gg \gamma_2 \gamma_2^\prime \bar{n}_2$ (where we have exploited the fact that $ \Delta\bar\omega_2^{eff} \simeq \Delta\omega_2^{eff} \simeq \Delta\omega$).
This transition to synchronization is consistent with the theoretical analysis made in Refs. \cite{Holmes:2012aa,Li2020}, which, in the regime of not large driving power studied here, predicts an onset of synchronization with very different limit cycle amplitudes, even in the presence of thermal noise. In the four--dimensional phase space of the mechanical oscillators it manifests itself via a Neimark-Sacker bifurcation corresponding to the birth of a stable torus around the existing limit cycle \cite{Balanov2008}.

We corroborate such analysis by considering the experimental time traces shown in Fig.~\ref{fig:Fig_20200110_Set3_Brownian_2D_Volt}.
The voltage ${\rm V}_{{\rm x}j}$ and ${\rm V}_{{\rm y}j}$ [panel a) and b)], are the slowly varying quadratures of the voltage signal, integrated over a bandwidth of \SIrange{70}{150}{\hertz} around the mechanical frequencies $\omega_1$, and $\omega_2$, as a function of time. In panel c) and d)  are reported the associated phase--space distributions. Panel e) and f) show the phase--space distributions of the calibration tone before and after the pump is turned on, confirming that the calibration tone is not appreciably affected.

\begin{figure}[h!]
\begin{center}
   {\includegraphics[height=.375\textwidth]{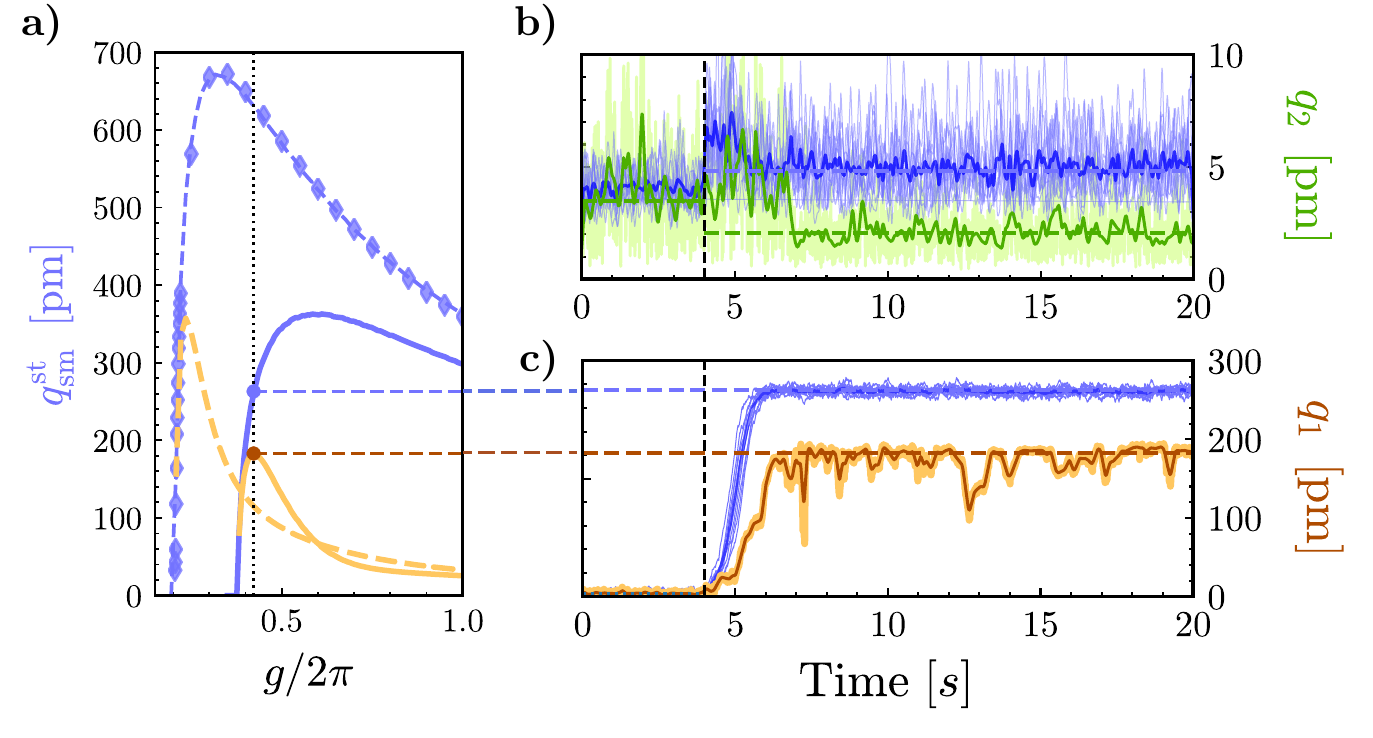}}
 \caption{
 		{\bf a)} Observed (orange curve) and effective (blue curve), steady--state amplitudes of the first mode, $q_{\rm sm}^{\rm st}$, as a function of the single--photon optomechanical coupling rate of the first mode $g \equiv g_1$. Solid, and dashed curves represent the solution of eq.~(\ref{eq:first-order finnal2}) for two pump settings: $P_{probe} = \SI{5.9}{\micro\watt}$, $P_{pump} = \SI{4.25}{\micro\watt}$, and $P_{probe} = \SI{16}{\micro\watt}$, $P_{pump} = \SI{18.7}{\micro\watt}$, respectively. 			Filled diamonds, which confirm the results obtained by finding the first zero of eq.~(\ref{eq:first-order finnal2}) for the second power setting, represent the effective steady--state amplitudes obtained as the mean of the amplitudes after the oscillator has reached the limit cycle, determined by integrating eq.~(\ref{eqampliappr1}) [and eq.~(\ref{eqampliappr2})].
		Filled circles represent the expected amplitudes for the experimental parameters given in Tab.~\ref{tb:tab1}: $q^{\rm st}_{\rm sm} \equiv q_1^{\rm st}= \SI{263.0}{\pico\meter}$, and $q^{\rm ob}_{1} = \SI{183.0}{\pico\meter}$.
		{\bf b)} Observed displacement amplitudes, $q_2$, as a function of time for the fundamental mode of the second oscillator, and {\bf c)} for the first oscillator, $q_1$, which reaches a limit cycle.
		Orange and green curves indicate experimental data, and darker curves a convolution over 200 points.
		The measured steady--state values are $q_{1} \simeq \SI{184}{\pico\meter}$, and $q^{\rm st}_{2} \simeq \SI{2.1}{\pico\meter}$, to be compared with the expected values $\SI{183.0}{\pico\meter}$, and $\SI{2.0}{\pico\meter}$, respectively.
		The 10 blue trajectories in panel c) represent the dynamics obtained by integrating eq.~(\ref{eq:first-order finnal0}) [and eq.~(\ref{eq:first-order finnal1})].
		The dashed horizontal lines in panels b), and c), indicate the expected values obtained by using eq.~(\ref{eq:gamma2_eff}), and derived in panel a), respectively.
	}
\label{fig:Exp_Theory_ALL}
\end{center}
\end{figure}

By means of the calibration tone~\cite{Gorodetsky:2010uq}, firstly, we determine the displacement amplitudes of the two oscillators, shown in Fig.~\ref{fig:Exp_Theory_ALL}b), and Fig.~\ref{fig:Exp_Theory_ALL}c), which, before the pump beam is turned on ($t <\SI{4}{\second}$), show higher values than the thermal ones. This is ascribed to a slightly blue--detuning of the probe beam. We evaluate such detuning observing that, for the second mode, green curve in panel b), the calibrated measured position standard deviation $\Delta q^{\Delta}_2 \simeq \SI{3.50}{\pico\meter}$, while the estimated thermal position standard deviation is $\Delta q_2^{\rm th} \simeq \SI{3.32}{\pico\meter}$, so that
\begin{eqnarray}\label{Dq_Dqth}
	\frac{\Delta q^{\Delta}_2}{\Delta q_2^{\rm th}} = \frac{1}{\sqrt{1+C(\Delta)}}\simeq \num{1.054}
	\,,
\end{eqnarray}
where, for almost resonant field~\cite{Aspelmeyer2014},
\begin{eqnarray}\label{C_Delta}
	C(\Delta) \sim -\frac{2g_j^2 E^2}{\gamma \kappa}\frac{4\omega_j}{(\kappa^2 + \omega_j^2)^2}\Delta
	\,.
\end{eqnarray}
This fact allows us to estimate a small blue detuning of the probe $\Delta_2 \simeq 2\pi\cdot\SI{3.9}{\kilo\hertz}$, which is the value provided in Table~\ref{tb:tab1}.

Finally we analyse the measured mechanical amplitudes after the pump beam has been turned on ($t>\SI{4}{\second}$): the amplitude of the first oscillator increases, while the measured amplitude of the second one reduces below the thermal value.
For the first oscillator, the observed steady--state limit cycle displacement amplitude is $q^{\rm ob}_1\simeq \SI{184}{\pico\meter}$, orange curve in Fig.~\ref{fig:Exp_Theory_ALL}c).
Such value agrees very well with the expected one, shown as dark--orange filled circle in panel a) of Fig.~\ref{fig:Exp_Theory_ALL}.
Panel a) represents the effective steady--state mechanical amplitudes obtained as solution of eq.~(\ref{eq:first-order finnal2}) (blue curves), and the observed one (orange curves), that is, reduced by the correction factor reported in Fig.~\ref{fig:Ratio_ALL}, and calculated in Section~\ref{sec:Probe analysis}.
For a given set of parameters we observe that both the effective and observed steady--state mechanical amplitudes reach a maximum as a function of $g$, and then decrease.
This result is confirmed by integration of eq.~(\ref{eq:first-order finnal0}) [and eq.~(\ref{eq:first-order finnal1})], including the noise contributions, and finding the steady--state as mean amplitude after the oscillator has reached the limit cycle, values which are reported as diamond symbols in Fig.~\ref{fig:Exp_Theory_ALL}a). The effective steady--state displacement amplitude of the first oscillator, $q^{\rm st}_1 = 2|A_1|x_{\rm zpf}  \simeq \SI{262}{\pico\meter}$, is also confirmed by the 10 blue trajectories simulated with the parameters of Table~\ref{tb:tab1}, and reported in Fig.~\ref{fig:Exp_Theory_ALL}c).
We note that even the slope of the trajectories follows with accuracy the measured one, implying that our approach in terms of slowly--varying complex amplitudes of the two oscillators, is effective, and able to grasp all the features of the non--linear dynamics.

Lastly, we observe that even the dynamics of the second mode is very well described by our model.
In fact, it is evident that after the pump is turned on, the behaviour of the observed $q_2$ [green curve in  Fig.~\ref{fig:Exp_Theory_ALL}b)] follows the dynamics of the effective mechanical displacement [blue trajectories in Fig.~\ref{fig:Exp_Theory_ALL}b)] only until $q_1$ reaches the limit cycle (after \SI{7}{\second}), and since then the observed displacement differs from the effective one.
An estimation of the effective steady--state amplitude of the second contribution in eq.~(\ref{a2stat}) is provided by $\Delta q_2^{\rm th}\sqrt{\gamma_2/\gamma_2^{eff}}$, with
\begin{eqnarray}\label{eq:gamma2_eff}
	\frac{\gamma_2^{eff}}{\gamma_2} = 1 + \frac{\text{Im}[d_2(I_1^{\rm st})]}{\gamma_2}
		= 1 + \frac{g_2^2E_1^2}{\gamma_2I_1g_1}\text{Im}\left[\Sigma_1 + \eta\,\Sigma_2\right]
		= 1 - \frac{\gamma_1g_2^2}{\gamma_2g_1^2}\,.
\end{eqnarray}
We note that, in our case, the effective amplitude is larger than the thermal one by a factor $\sqrt{\gamma_2/\gamma_2^{eff}}\simeq 1.38$, that is, there is a small effective driving, although not enough for the appearance of a limit cycle. Also, from the correction factor $\calN_2 \simeq \num{0.42}$, we estimate an observed displacement of $q_2^{\rm ob} = \Delta q_2^{\rm \Delta} \cdot\sqrt{\gamma_2/\gamma_2^{eff}}\cdot \calN_2 \simeq \SI{3.50}{\pico\meter}\cdot\num{1.38}\cdot\num{.42} \simeq \SI{2.0}{\pico\meter}$, in great agreement with  the measured value of $\SI{2.1}{\pico\meter}$.
In conclusion, we observe that even the small effective amplitude displacement of the second oscillator is strongly affected by the non--linear cavity dynamics, exhibiting a fictitious cooling effect, which is instead only a manifestation of detection in this nonlinear regime. This is a somehow unexpected effect of the nonlinear regime in which our system operates. As soon as the amplitude of one of the resonators yields a frequency modulation larger than the cavity linewidth, \textit{all} the optically detected motional amplitudes are nonlinearly modified and appropriate calibration factors ${\cal N}_j$ must be considered. This occurs also to the unexcited resonator whose motional amplitude corresponds to a cavity frequency modulation much smaller than the cavity linewidth.

\section{Conclusion}

We have presented a detailed experimental analysis of the dynamics of the multimode optomechanical system introduced in Ref.~\cite{Piergentili:2018aa}, formed by a sandwich of two membrane mechanical resonators placed within a high-finesse cavity, and interacting with a pump and a probe cavity mode. We have focused onto the non-linear regime where a blue-detuned pump drives one of the two oscillators into a self-sustained limit cycle. In the weak driving regime studied here, the system is in a pre-synchronized situation where the unexcited oscillator has a small, synchronized component at the frequency of the excited (master) oscillator, which is however dominated by the fluctuating thermal noise component. We find perfect agreement between the experimental results, the numerical simulations, and an analytical approach based on slowly-varying amplitude equations. This analytical study allows to derive a full and consistent description of the displacement detection by the probe beam in this non-linear regime, enabling the faithful detection of membrane displacements well above the usual sensing limit corresponding to the cavity linewidth. In this non-linear detection regime, both large and small amplitude resonator motion are transduced in a nontrivial way by the non-linear response of the optical probe beam.

\section*{Acknowledgments}
We acknowledge the support of the European Union Horizon 2020 Programme for Research and Innovation through the Project No. 732894 (FET Proactive HOT) and the Project QuaSeRT funded by the QuantERA ERA-NET Cofund in Quantum Technologies. P. Piergentili acknowledges support from the European Union's Horizon 2020 Programme for Research and Innovation under grant agreement No. 722923 (Marie Curie ETN - OMT).
\\
P.~P. and W.~L. contributed equally to this work.

\end{document}